\documentclass[]{aastex631}
\usepackage{wrapfig}

\received{March 8, 2022}
\accepted{September 21, 2022}

\submitjournal{ApJ}

\shorttitle{Statistics of Post-Flare Coronal Rain}
\shortauthors{Mason \& Kniezewski}
\graphicspath{{./}{figures/}}

\begin{document}

\title{To Rain or Not to Rain: Correlating GOES Flare Class and Coronal Rain Statistics}

\correspondingauthor{E I Mason}
\email{emason@presci.com}

\author[0000-0002-8767-7182]{E I Mason}
\affiliation{Predictive Science Inc. \\
9990 Mesa Rim Rd, Suite 170 \\
San Diego, CA 92121, USA}

\author[0000-0003-3740-9240]{K Kniezewski}
\affiliation{United States Naval Academy \\
121 Blake Road \\
Annapolis, MD 21402, USA}

\begin{abstract}
Post-flare arcades are well-known components of solar flare evolution, which have been observed for several decades. Coronal rain, cascades of catastrophically-cooled plasma, outline the loops and provide eye-catching evidence of the recent flare. These events are acknowledged to be common, but the scientific literature does not include any statistical overview documenting just how common the phenomenon actually is. This study reviews Solar Dynamics Observatory Atmospheric Imaging Assembly (SDO AIA) observations of 241 flares collected from the Space Weather Prediction Center (SWPC) database between 2011 and 2018. The flares cover the entire strength range of the C, M, and X GOES classes, and are distributed evenly across the SDO-observed majority of Solar Cycle 24. We find that post-flare arcade rain occurs for nearly all X and most M-class flares, but that it tapers off rapidly within C-class flares. There appears to be a cut-off point around C5, below which the occurrence of post-flare arcade rain drops significantly. There is also a general positive correlation between GOES class and the average duration of post-flare rain events. Post-flare arcade rain events in X- and M-class flares appear to track with the sunspot number, providing a potential new tool for estimating, if not predicting, solar cycle strength. Furthermore, arcades are observed to persist for up to several days after the originating flare, transitioning from hosting post-flare rain to typical quiescent active region condensations. These results open up further avenues for future research, including new methods to estimate energy deposition and to gain greater insight into steady active region heating.
\end{abstract}

\section{Introduction} \label{sec:intro}

This paper addresses coronal rain events in solar flares, investigating the frequency with which such events occur, their duration, and other salient details. We correlate these statistics to overarching open questions concerning the Sun: flare evolution, coronal mass ejection (CME) creation, solar cycle characteristics, and localized coronal heating. 

White-light study of solar flares began in 1859 with the independent but concurrent observations of \cite{Carrington1859} and \cite{Hodgson1859}, and investigation of these explosive events spread to both longer and shorter wavelengths -- sometimes intentionally, sometimes serendipitously -- as technology advanced. The current model for flares, known as the standard model and succinctly summarized by \cite{Holman2016}, holds that magnetic reconnection across field lines arrayed along a polarity inversion line releases energy, heating the plasma to temperatures in excess of 30 MK and accelerating particles along the newly-rearranged magnetic field. Flares are often cited as sources of solar energetic particles (e.g., \cite{Reames1990}, \cite{Cliver2012}, \cite{Reames2017}), radio bursts (for starters, see \cite{Wild1972}, \cite{Aschwanden1985}, \cite{Nelson1985}, and \cite{Cane2002}), and CMEs (\cite{Feynman1994}, \cite{Svestka2001}, \cite{Andrews2003}, \cite{Yashiro2005}, \cite{Cane2006}, among many others).

According to the flare Standard Model (\cite{Holman2016}), shearing along a polarity inversion line (PIL) elongates field lines; eventually, some form of disturbance can cause these lines to distort, triggering reconnection along the PIL. This builds up a flux rope (which carries twist, translated upward into the corona from the shearing) above an arcade of loops that are arrayed roughly perpendicular with respect to the PIL. If the flare is eruptive, the flux rope is then ejected into the outer corona and the solar wind as a CME (\cite{Lin2000}, \cite{Mittal2010}); otherwise, it is considered a confined flare. The conditions which determine whether or not a flare is confined are not well-known (\cite{Sun2015}, \cite{Zuccarello2017}, \cite{Mason2021}). These events carry magnetic field as well as plasma into the heliosphere, and can strongly affect systems both on the Earth and in space. Early observations tied their existence almost solely to solar flares, but it was subsequently found that CMEs also frequently form from more quiescent events like prominence eruptions (\cite{Gosling1993}). However, strong (and particularly long-duration) flares are well-correlated with CMEs, as discussed in \cite{Sheeley1983} and \cite{Yashiro2005}.

The arcade loops that form below the flux rope are filled with extremely hot, dense plasma through electron beam heating, in a process termed "chromospheric evaporation". This results in a loop which is severely outside the normal coronal conditions, resulting in thermal instability (\cite{Parker1953}, \cite{Priest1978}, \cite{Antiochos1980}, and \cite{Klimchuk2019}, among many others). The temperature and density in the loop are unsupported by the energy being added by whatever ambient background or footpoint heating is occurring in the arcade loop, so the loop begins radiating rapidly. Here we reach the end of what is well-understood about flare arcades: if the temperature profile of the loop was uniform, we would expect the arcade loop to cool and evacuate in its entirety, as was found by \cite{Reep2020} for loops heated purely by the electron beams. Contrary to this, the recent work of \citet{Ruan2021} successfully reproduced post-flare rain with a 2.5D MHD simulation without even including beams. That model gets sufficient heating and condensations simply from the magnetic energy conversion during flare reconnection. Observations show that individual condensations occur within arcade loops; this implies instead that a localized temperature inversion is critical to the post-flare arcade rain formation mechanism, to seed nascent rain blobs.

In this work, we focus on these diminutive condensations. It is important to point out and clearly define the phenomenon which we study here, because it delineates an important distinction for observations. Any typical active region will host significant numbers of coronal condensations on its loops during its lifetime, and these may increase in number in the time after the impulsive phase of a flare. However, since our goal in this paper is to quantify that rain which is an immediate effect of flare reconnection, the most natural observational selection was to only consider condensations formed within the flare arcade. One of the main findings of this study, however, was the discovery that \emph{the arcades themselves can often become simply other coronal loops that host quiescent rain, given enough time}: this frequently occurs with medium-sized flares, where the arcade grows to a point but does not vanish. Since the rain which can occur continuously in these loops several days after the flare is almost certainly not causally linked to the flare, this rain also cannot be considered. Therefore, in our observations we also needed a marker for the end of the rain duration. We define it to be at the point that the raining arcade loops ceased to increase in height (best for off-limb flares), or when the flare ribbons stopped evolving (best for on-disk flares). Both of these measures -- which are two observables of the same physical process -- signal that the flare reconnection has ceased and that further rain on the same loops is likely due to another heating source. Using this narrow definition, we consider only the rain directly tied to flare reconnection, producing results that are the most relevant to flare dynamics. This is understood by our use of the common terms \emph{post-flare rain} or \emph{post-flare arcade rain} throughout this paper.

One of the greatest benefits of the Solar Dynamics Observatory Atmospheric Imaging Assembly \citep[SDO AIA;][]{Pesnell2011,Lemen2011} has been the sheer number of events which have been captured in detail by the near-constant view across a broad range of extreme ultraviolet (EUV) passbands. Much of the flare literature revolves around in-depth case studies of individual flares, or small collections of similar flares. Here we conduct an assessment of post-flare rain in a large pool of events. Rain has been reported for years as a feature of strong flares, but the frequency and statistical properties of it has not, to our knowledge, ever been quantified. While not the brightest signal of the elusive magnetic dynamics involved in flares, post-flare rain provides a critical reservoir which we can use to measure how much of the flare's energy is actually deposited in the plasma still trapped in the corona. Energy which is radiated away as the condensations form is energy that is no longer contained in the magnetic field, powering a CME or an SEP event (\cite{Reames2013}, \cite{Drake2012}). We review all of the X-class flares that occurred between 2011 and 2019, as well as a large, representative sample of M- and C-class flares from the same period (we detail the collection and analysis of these in Section \ref{s:data}). This survey allows us to build a strong basis for general statements about energy budgeting and magnetic field dynamics in more energetic flares.

\section{Data Collection and Analysis}\label{s:data} 

\subsection{Data Selection}

We queried the Heliophysics Events Knowledgebase (HEK; accessible at \url{https://www.lmsal.com/hek/index.html}) for flares detected by the GOES mission between 2011 January 1 and 2019 December 31. We selected this date range to cover most of Solar Cycle 24 for which we have routine daily SDO coverage. Solar Cycle 24 began in December 2008, but SDO finished commissioning and began its main science mission in May 2010. During this time period, there were 49 X-class flares,  732 M-class flares, and 7681 C-class flares. Since this study was conducted manually, a subset of the M and C-class flares were selected; all of the X-class flares were analyzed. We chose flares to create a pool that covered roughly equal numbers of flares in all relevant characteristics: class (i.e., M), intra-class magnitude (i.e., M5), and distribution across the solar cycle (equal numbers from each year covered by the study, when possible). All the flares in the data set were selected for their adherence to these characteristics, and prior to analysis; a flare was not viewed until it had been selected for the study. If enough flares in a category were unable to be analyzed due to data dropouts, eclipsing of SDO, or similar reasons such that it would skew the distribution of the final data set, more flares were chosen from the list.

The final selection included all 49 X-class flares, as well as 78 M-class and 118 C-class flares. 5 of the X-class flares had data dropouts which precluded analysis, so the statistics here are on the 44 which had available AIA data. A few other events had truncated rain durations because of data dropouts or subsequent flares in the same location; these are all noted as such in the full dataset, available on Zenodo at \url{https://doi.org/10.5281/zenodo.7028211}.
From the GOES flare list, we obtained the flare start, peak, and end times; the location on the Sun in heliographic coordinates; and the source NOAA active region designation.

\subsection{Observational Analysis and Challenges}

\begin{wrapfigure}[23]{l}{0.5\textwidth}
\vspace{-\intextsep}
\centering
    \includegraphics[width=1.\linewidth,clip]{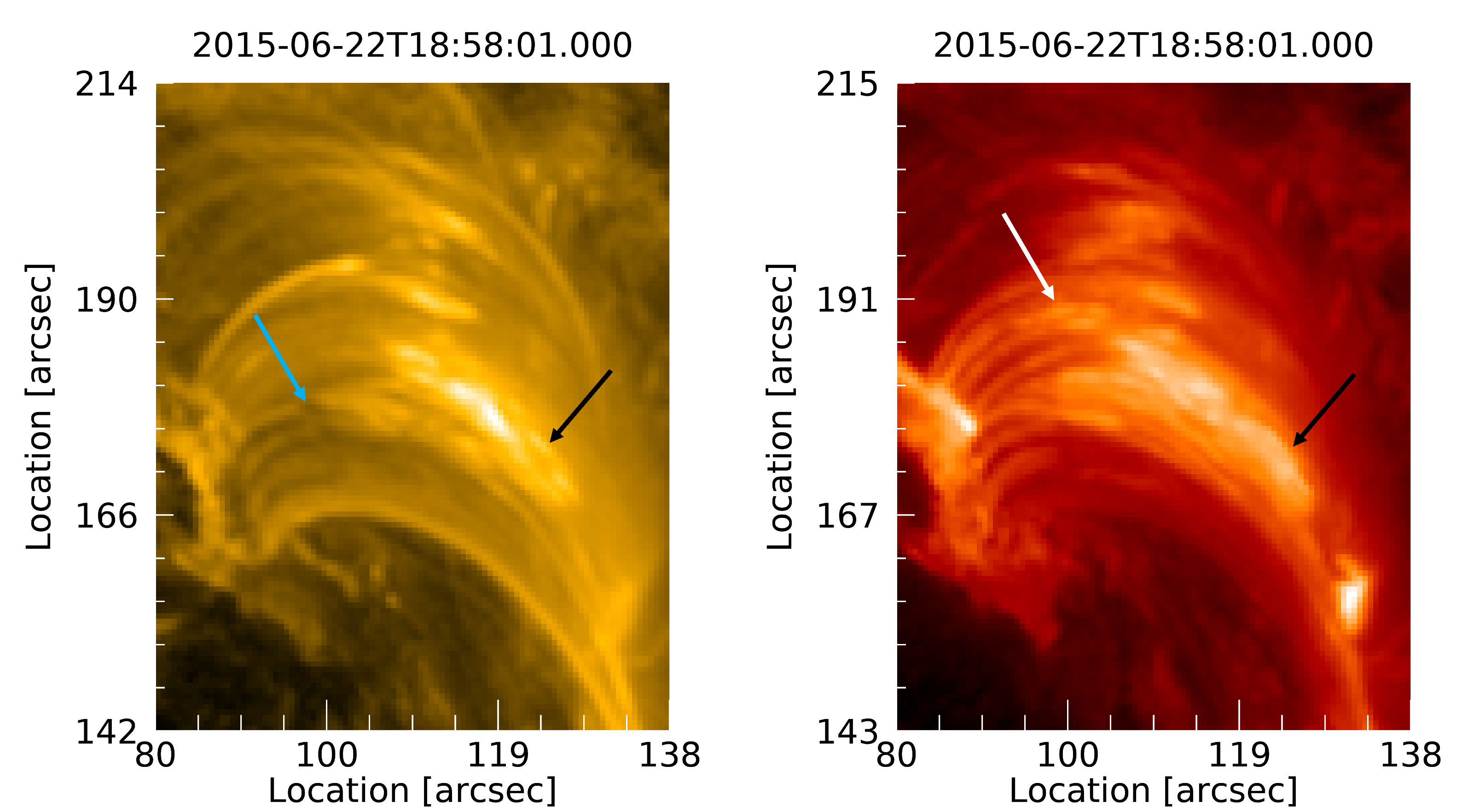}
              \caption{Left: Screen capture of animation (Movie 1, 5 seconds in duration, available online showing 108 minutes of flare arcade evolution) showing the flare arcades from an M6.5 flare in SDO AIA 171 \AA. An example cool condensation is indicated in absorption with the blue arrow, and in emission with a black arrow. Right: The same flare as seen in Movie 1 in 304 \AA, with adjusted scaling to resolve the condensations. An example post-flare arcade rain blob is indicated in emission with a white arrow, which is not seen in the left-hand panel, while the black arrow corresponds to an analogous condensation signature as that indicated by the black arrow in 171.}
              \label{F1}
\end{wrapfigure}

The analysis itself was primarily conducted via visual inspection of SDO AIA data. Movies were generated with JHelioviewer (\cite{Muller2017}), showing eight hours of data at a time beginning with the start time of the flare, and 171 and 304 {\AA} data were evaluated for signs of cool condensations (see Figure \ref{F1} for a still-frame of one such movie) within flare arcades. Observationally, arcade loops appear as relatively short, bright loops with footpoints rooted at (indeed, defining) the flare ribbons. They generally form perpendicular to the solar surface, and are almost always symmetric about their apex. These were the observational hallmarks we used, which include characteristics that make them readily identifiable both on-disk and off-limb. 

When post-flare rain was present up until the end of the 8-hour movie, the inspection time period was extended (this occurred primarily for X-class flares, as can be seen from Figure \ref{F2}). 304 {\AA} is the coolest main SDO EUV channel, and the one most commonly used to observe coronal rain in emission; we chose 171 {\AA} for its cool, well-defined loop structures in active regions, where condensations can often appear clearly outlined as both emission and absorption features. Hotter lines also sometimes show rain in absorption, but the attendant diffuse glow from the active region obfuscates individual loops, and so these were not widely used in our analysis. Once post-flare arcade rain was established to be observed in 304, the duration of the event was observed and recorded, in accordance with the definition we outlined in Section \ref{sec:intro}.

There are, of course, pitfalls to attempting to observe small, cool structures in 304, particularly when the flare occurs on disk. One of these challenges is that 304 frequently saturates early in a flare, and some detail is lost. As can be seen from Figure \ref{F1}, we adjusted the maximum and minimum values for the color table from the default scaling that is commonly applied to AIA data; while this did not entirely eliminate the early saturation during the impulsive phase, it did frequently resolve structure significantly earlier than would normally be observed with the standard scaling rules. The early saturation is a well-known problem which cannot be overcome without introducing another instrument, which would have significantly restricted our pool of events.

Another common observational challenge relates to identifying cool condensations when the flare occurred on-disk. Since 304 {\AA} observations are dominated on-disk by chromospheric and transition region emission, rain can be difficult to pick out. However, this problem affects quiescent rain more than post-flare arcade rain; when observing the flares, the activated source region was bright enough that the cool blobs were easily viewed the majority of the time. For cases where the observation was somewhat ambiguous, we searched for transient dark structures (e.g., blobs, streaks, or lines) running perpendicular to the flare ribbons in both 171 and 304. Since 171 is the next coolest channel to 304, it was often possible to see a brightening form and then dim in 171, followed by a cospatial signal in 304. For the handful of highly ambiguous cases, we used running difference movies to capture precipitating condensation motions that were otherwise masked by more random activity in the region.

The last but most pervasive challenge when observing medium-to-large flares did not involve problems with observing the condensations themselves, but rather how to define when flare-related rain transitioned to the typical active region quiescent rain. For many of the flares analyzed here, the arcade loops did not disappear after the flare; they either began raining again, or continued raining without pause for tens of hours after all other observational signatures of the flare had ended. An example of this can be seen in the movie associated with Figure \ref{F1}. As we discussed in Section \ref{sec:intro}, the most logical, physically-motivated dividing line was the point at which the arcade reached its apparent maximum height or the point when the flare ribbons ceased evolving (these being physically equivalent), which both imply that the flare reconnection forming the arcade has ended. Within a few minutes, the last arcade-coupled rain should fall, and subsequent rain should be motivated by more quiescent loop conditions.

Due to the limitations inherent in remote EUV observations with a resolution lower than that of the smallest condensations, there are likely to be smaller rain events that are unresolved or lost in the background signal. There are also line-of-sight constraints which make it difficult to determine when the coronal rain has reached the chromosphere. None of the uncertainties presented in this section are unique to this study, but are indeed shared by all remote solar observations. While it is possible to wait for a prime alignment to observe when studying long-duration structures, the impulsive nature of solar flares often means taking measurements at imperfect angles. This study, therefore, represents a uniformly-analyzed body of flares which were chosen for the best possible statistical coverage, whose condensations were identified by eye and were categorized with a particularly restrictive definition in order to provide the most physically meaningful results.

We also determined the eruptivity of each solar flare event (whether or not a CME was released) through HEK  following flare events of interest. HEK was queried for 1-hour-long spans beginning at the flare start time, and we then assessed the returned CME or CMEs to determine whether the location correlated with the flare within the respective time frame. CME candidates for these flares were visually identified and confirmed through the SOHO LASCO CME Catalog.\footnote{This CME catalog is generated and maintained at the CDAW Data Center by NASA and The Catholic University of America in cooperation with the Naval Research Laboratory. SOHO is a project of international cooperation between ESA and NASA.}

\section{Statistical Results}\label{s:stats} 

\begin{figure}[ht]    
   \centerline{\hspace*{0.01\textwidth}
               \includegraphics[width=0.5\textwidth,clip=]{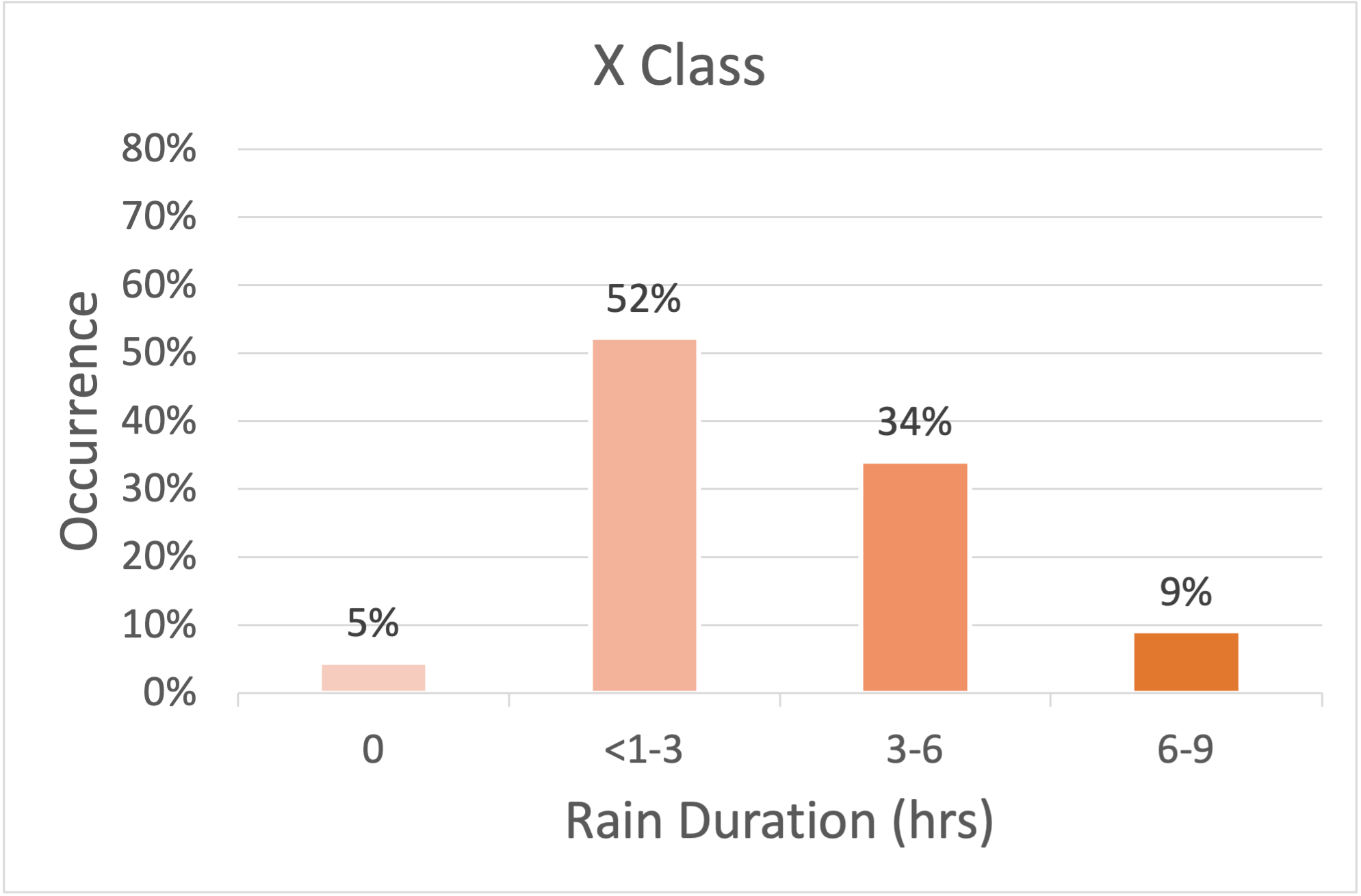}
               \hspace*{0.01\textwidth}
               \includegraphics[width=0.5\textwidth,clip=]{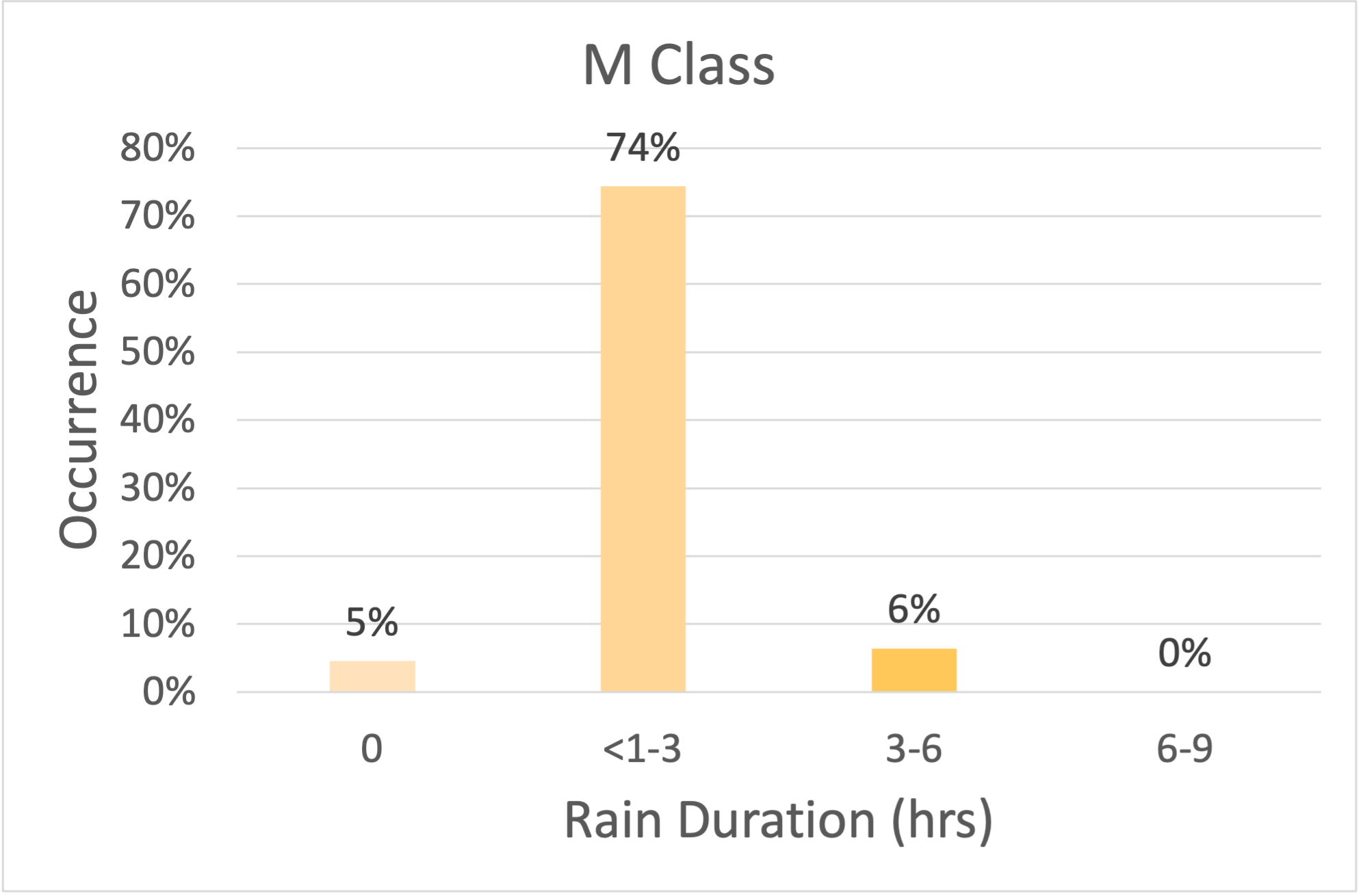}
              }
     \vspace{-0.35\textwidth}   
     \centerline{\Large \bf 
      \hspace{0.0 \textwidth}  \color{white}{(a)}
      \hspace{0.0\textwidth}  \color{white}{(b)}
         \hfill}
     \vspace{0.31\textwidth}    
   \centerline{\hspace*{0.01\textwidth}
               \includegraphics[width=0.5\textwidth,clip=]{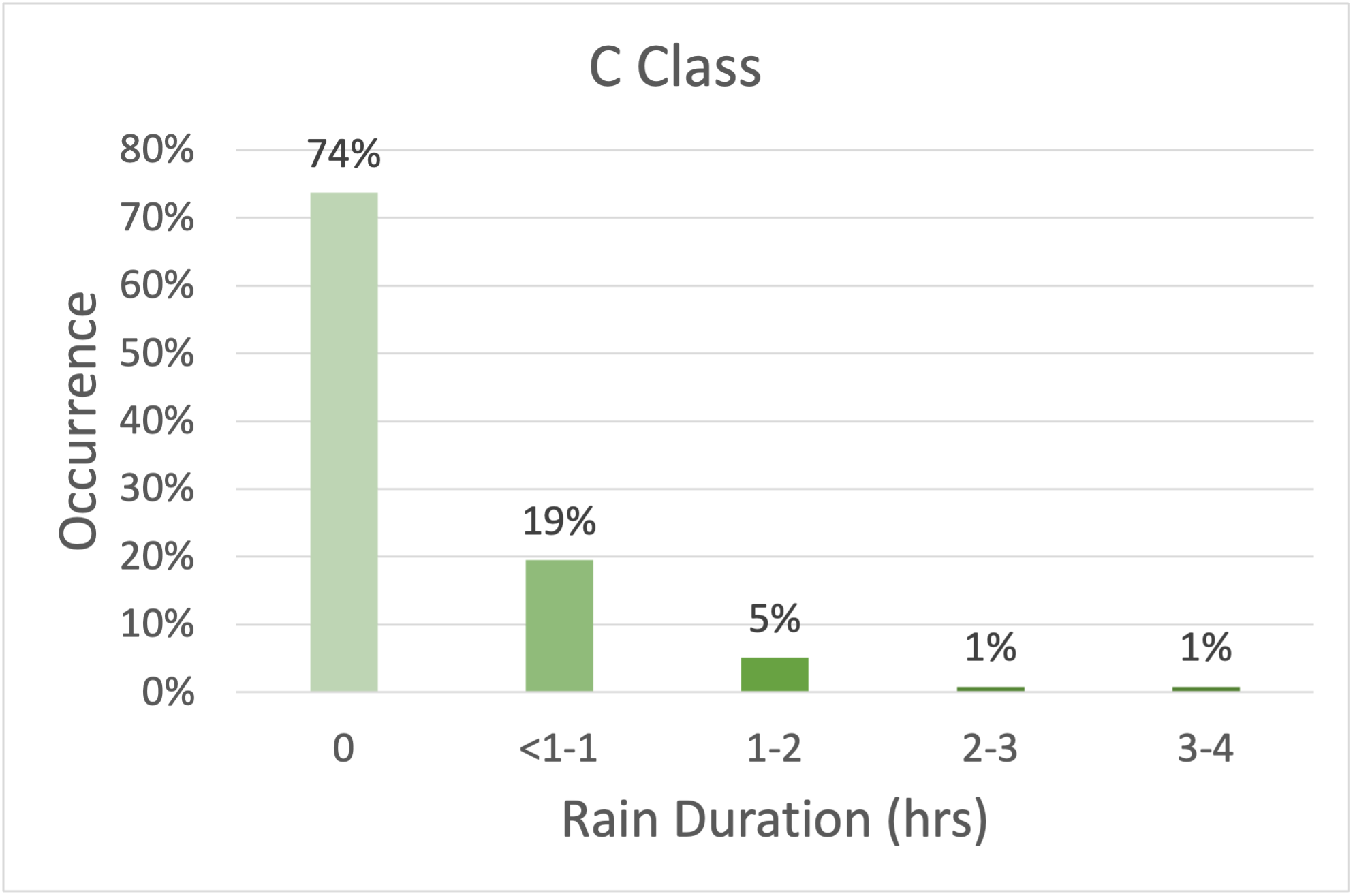}
               \hspace*{0.01\textwidth
               \includegraphics[width=0.5\textwidth,clip=]{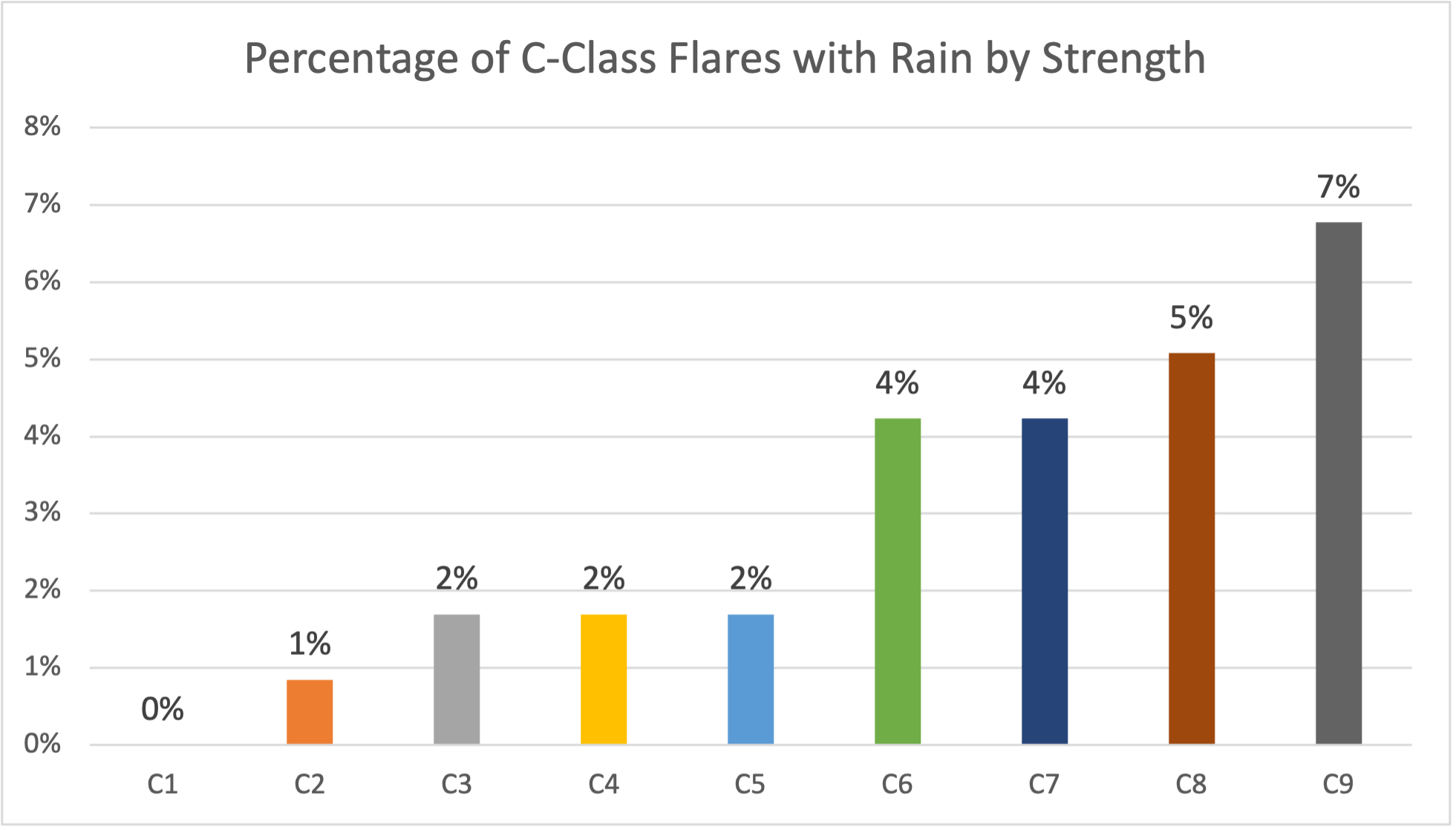}}
              }
     \vspace{-0.35\textwidth}   
     \centerline{\Large \bf     
      \hspace{0.0 \textwidth} \color{white}{(c)}
      \hspace{0.415\textwidth}  \color{white}{(d)}
         \hfill}
     \vspace{0.31\textwidth}    
              
\caption{Top left: histogram presenting the percentage frequency of X-class flares hosting post-flare arcade rain of duration ranging from under an hour to over 6} hours. Top right: analogous histogram for M-class flares. Bottom left: analogous histogram for C-class flares; please note that these are subdivided into shorter time bins only ranging between \textless1 hour and 4 hours, due to the absence of any longer post-flare arcade rain events detected in the survey. Bottom right: the percentages of flares which post-flare arcade rain by strength within the C-class designation, showing the strong change in post-flare arcade rain occurrence around C5.
   \label{F3}
   \end{figure}

Table \ref{T1} summarizes the proportions of flares which did and did not host post-flare arcade rain. Only two X-class flares out of the 44 did not produce coronal post-flare arcade rain; one of these, on 2012 March 5 at 02:30 UTC, was a large and unusual flare near the limb that simply did not form an arcade, but which was followed several hours later by a second flare with a classic arcade and several hours of post-flare rain (the second flare was not part of our analysis). This pair of flares is worthy of further study, to determine how they were related, and what magnetic changes would lead to their unique characteristics. The other (2013 November 8, 4:20 UTC) was in a small confined jet on the border of NOAA active region 11890, which did not exhibit an arcade large enough to be able to detect individual condensations. The preponderance of M-class flare arcades also exhibited post-flare arcade rain, although there is a slight drop of rain frequency between the X and M classes. C-class flares break this trend by showing a much lower propensity for condensation formation; only $26\%$ of these flares in the data set hosted post-flare arcade rain. Furthermore, this tendency showed an uneven distribution across the C flare magnitudes: the bottom-right panel of Figure \ref{F3} shows that there is significantly more rain observed in flares of intensity C5 or higher. As we discuss in the next section, this implies a minimum energy deposition required for condensations to form.

\begin{figure}[ht]    
   \centerline{\includegraphics[width=0.75\textwidth,clip=]{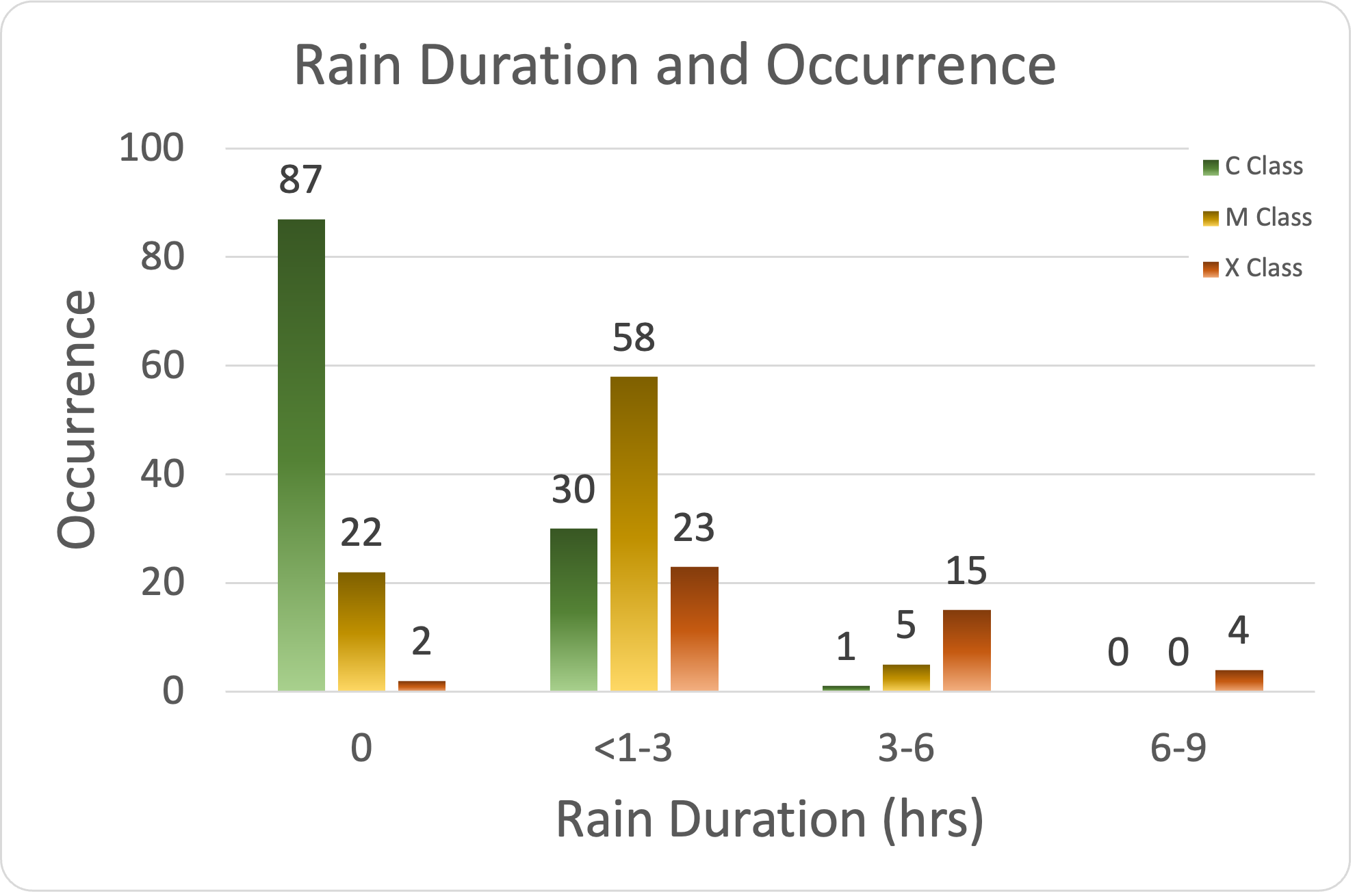}
              }
              \caption{Graph summarizing the duration of post-flare rain events by GOES class. X-class flares have a wide distribution of flare rain durations, while M-class flares' post-flare arcade rain most often lies in the \textless1-3 hour range. C-class post-flare arcade rain never exceeded 3 hours in duration in the sample studied here (the bins are not end-inclusive; for a more detailed breakdown of C-class rain times, see Figure \ref{F3}d).}
   \label{F2}
   \end{figure}

\setlength{\tabcolsep}{20pt}
\begin{table}[h!]
 \caption{Total Flares Analyzed by Post-Flare Arcade Rain Occurrence}
 \begin{tabular}{||c c c||} 
 \hline
  & \textit{Post-Flare Arcade Rain} & \textit{No Post-Flare Arcade Rain} \\ [0.5ex] 
 \hline
 \textit{X} & 42 (95\%) & 2 (5\%)\\ [0.5ex]
 \textit{M} & 56 (72\%) & 22 (28\%)\\ [0.5ex]
 \textit{C} & 31 (26\%) & 87 (74\%)\\ [0.5ex] 
 \textbf{Totals} & \textbf{129 (54\%)} & \textbf{111 (46\%)}\\ [0.5ex] 
 \hline
 \end{tabular}
 \label{T1}
\end{table}
Figure \ref{F2} shows the overall correlation between GOES class and the duration of post-flare rain events. X-class flares have a broad distribution between the \textless1 to 6 hour bins; the average rain duration was 2 hours 51 minutes. M-class flares show a strong peak of 58 in the \textless 1-3 hour range (average value 1 hour 14 minutes), which tails off precipitately for longer post-flare arcade rain events, while C-class flares have 87 events with no post-flare arcade rain and 30 with only short post-flare arcade rain ``storms" (average value 15 minutes); there was only one C-class flare with post-flare arcade rain slightly over 3 hours in length, and none longer. Figure \ref{F3} shows these correlations in more detail by graphing the time spans of post-flare arcade rain events for each class separately. The time bins there are uniform between X and M-classes, but are broken down into smaller time spans for the C-class flares. The correlation between flare class and post-flare arcade rain duration appears clearly even with a data set that spans a large representative sample of the M and C-class catalogs, so we consider the relationship strong and significant.

\begin{figure}[h]    
   \centerline{\includegraphics[width=0.7\textwidth,clip=]{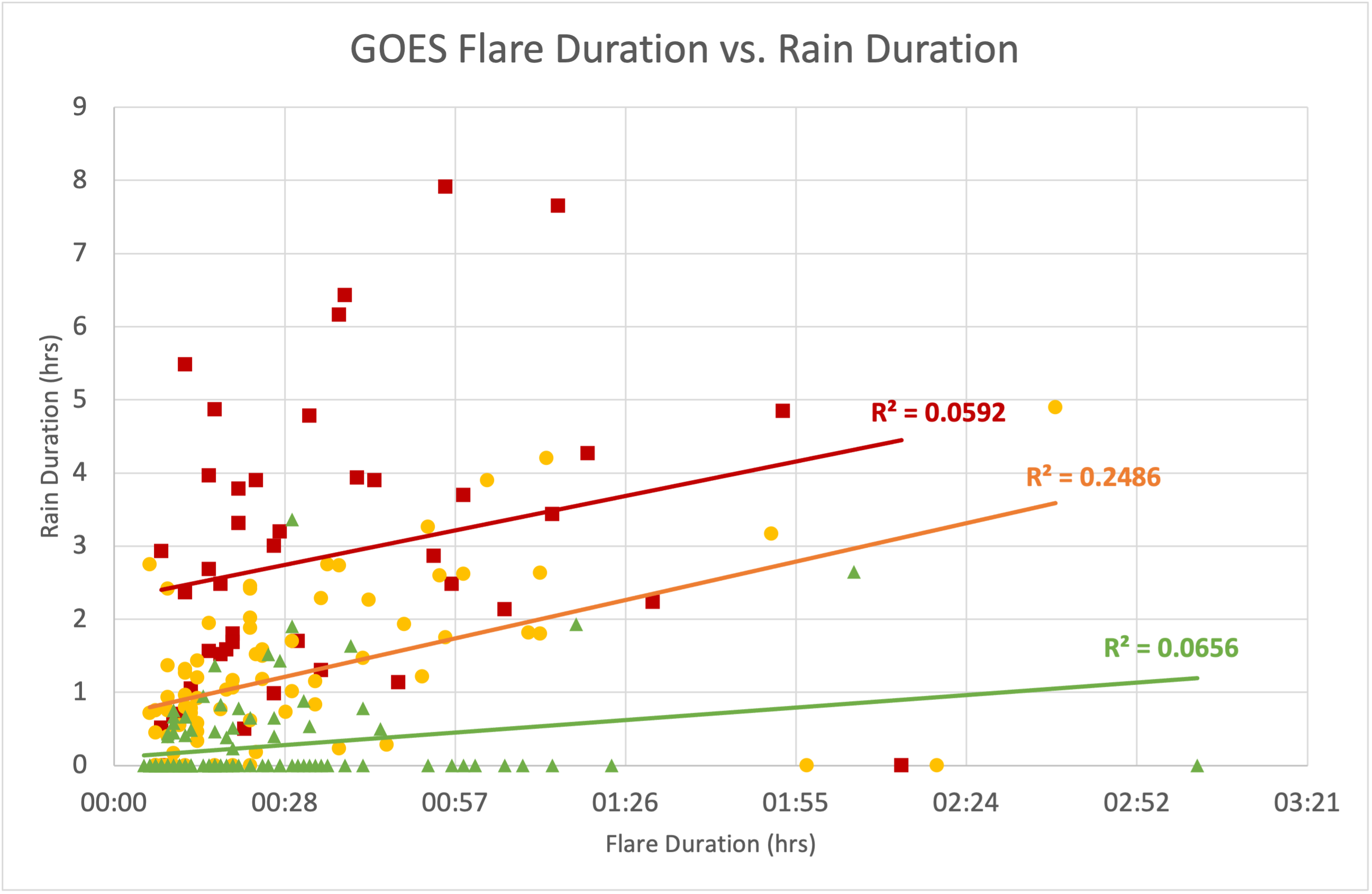}}
              \caption{Graph of flare duration, as determined by the GOES X-ray data, vs. post-flare arcade rain duration. Linear fits for each class show that all are generally a positive relationship, but none are statistically significant. M-class flares have the highest correlation, with an r-squared value of approximately 0.25.}
   \label{F4}
   \end{figure}

Given the very wide range of post-flare arcade rain durations which we found in the sample, we investigated whether there was a relationship between the length of the flare itself, as measured by the GOES X-ray intensity, and the duration of the subsequent post-flare arcade rain. Figure \ref{F4} shows the results of this investigation; we applied a linear regression to the full set of each class of raining flare and assessed the fit via the $r^2$ measure. None of the data shows a statistically significant relationship, although the M-class flares show the least-poor correlation, with an $r^2$ value of 0.2486. X-class flares tended to have long post-flare arcade rain events, regardless of the length of the flare, while C-class flares had a wide flare duration spread but very short rain durations. There has been some debate about the actual duration of C-class flares, and several background-subtraction methods have been put forward \citep[discussed recently in][]{Reep2019}. Given the very low rates of flare rain across all C-class flares, however, it is unlikely that background subtracting these low-energy flares would significantly improve the correlation.

  \begin{figure}[h]    
   \centerline{\includegraphics[width=.6\textwidth,clip=]{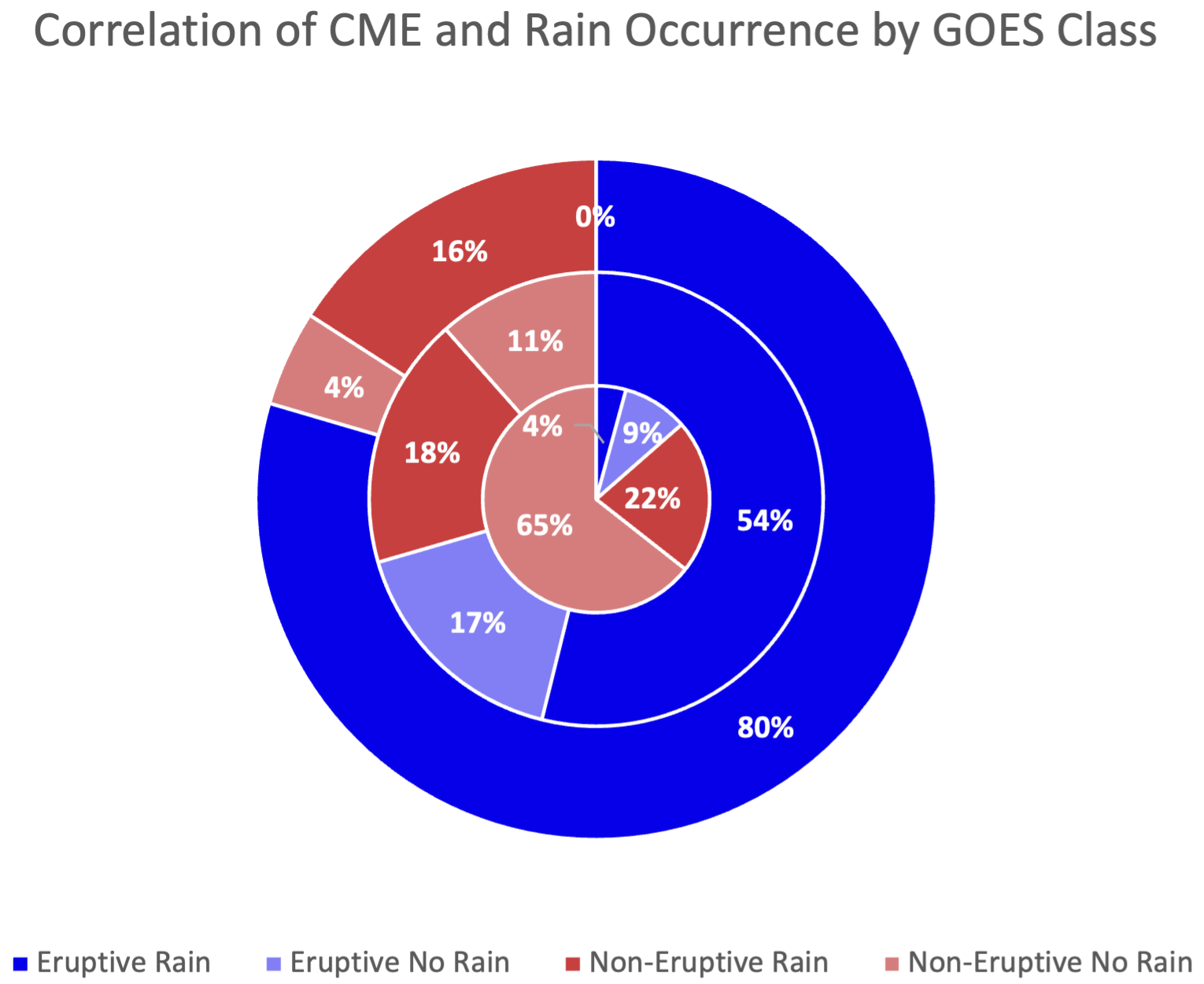}}
              \caption{This graph presents the proportions of X, M, and C class flares which are associated with CMEs and post-flare rain. There are few statistical conclusions that can be drawn between the raining and eruptive categories, due to the low rates of non-raining flares in X- and M-class flares, and the low rates of eruptive flares in C-class flares.}
   \label{F5}
   \end{figure}

  \begin{figure}[h]    
   \centerline{\hspace*{0.01\textwidth}
                \includegraphics[width=.7\textwidth,clip=]{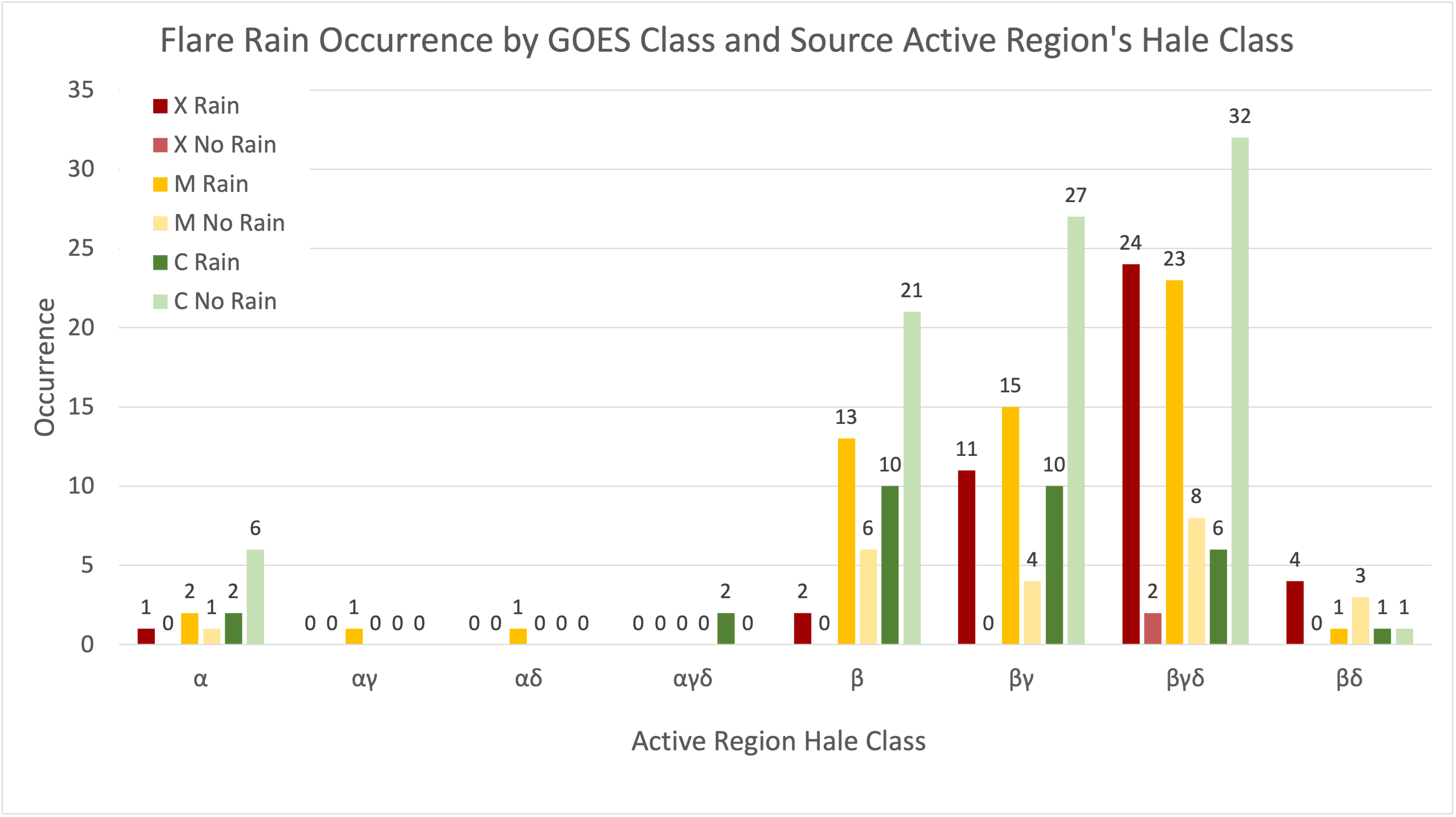}}
    \centerline{\hspace*{0.01\textwidth}
                \includegraphics[width=.7\textwidth,clip=]{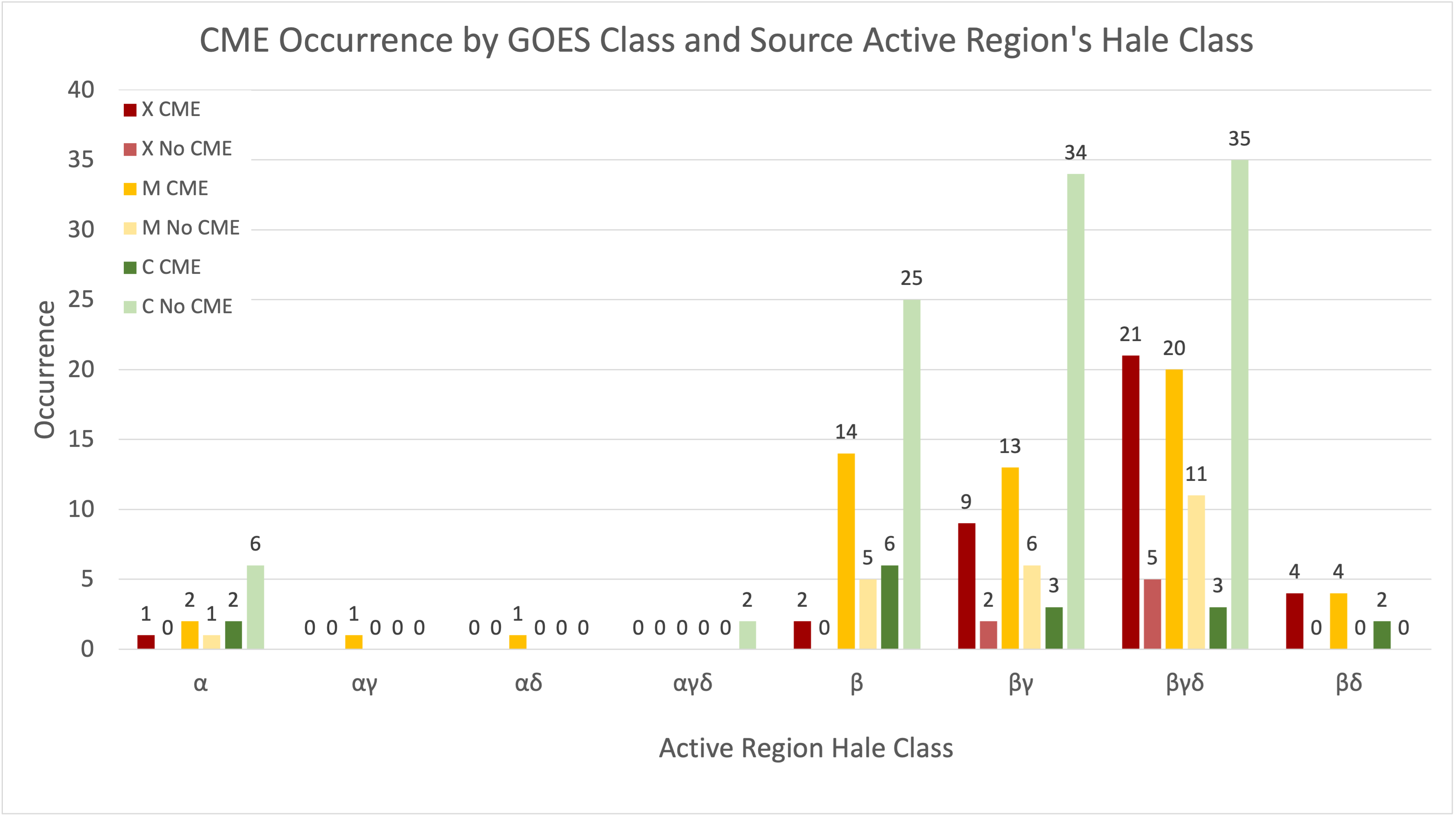}}
              \caption{Top: a histogram summarizing the relationship between post-flare arcade rain occurrence and the Hale class of the source active region (the bins along the x-axis), color-coded by GOES flare class. Bottom: an analogous histogram showing eruptivity of flares and the source region's Hale class. Both show strong preference for $\beta$ type active regions.}
  \label{F6}
   \end{figure}

  \begin{figure}[h]    
   \centerline{\hspace*{0.01\textwidth}
   \includegraphics[width=0.8\textwidth,clip=]{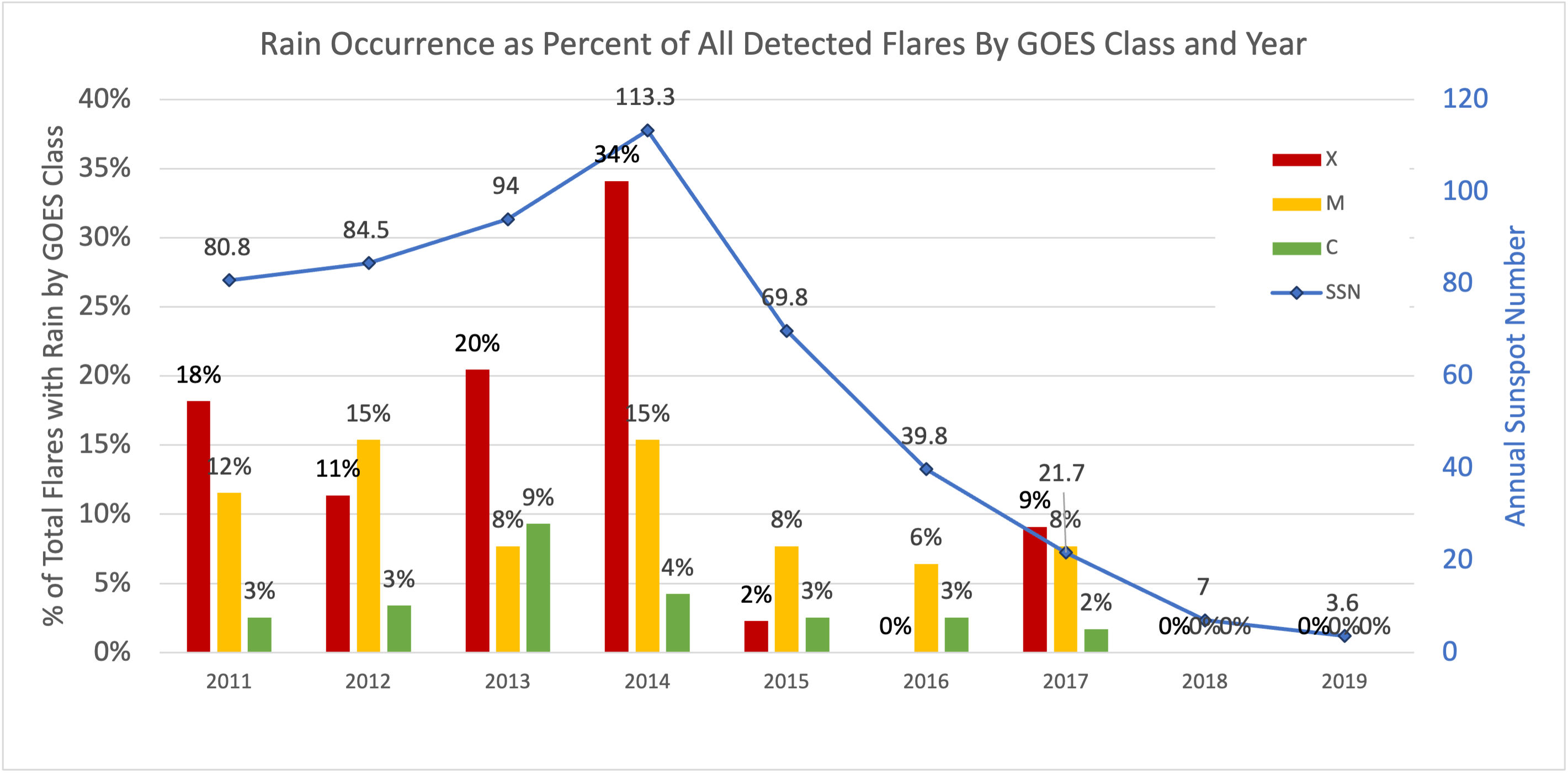}}
   \centerline{\hspace*{0.01\textwidth}\includegraphics[width=0.5\textwidth,clip=]{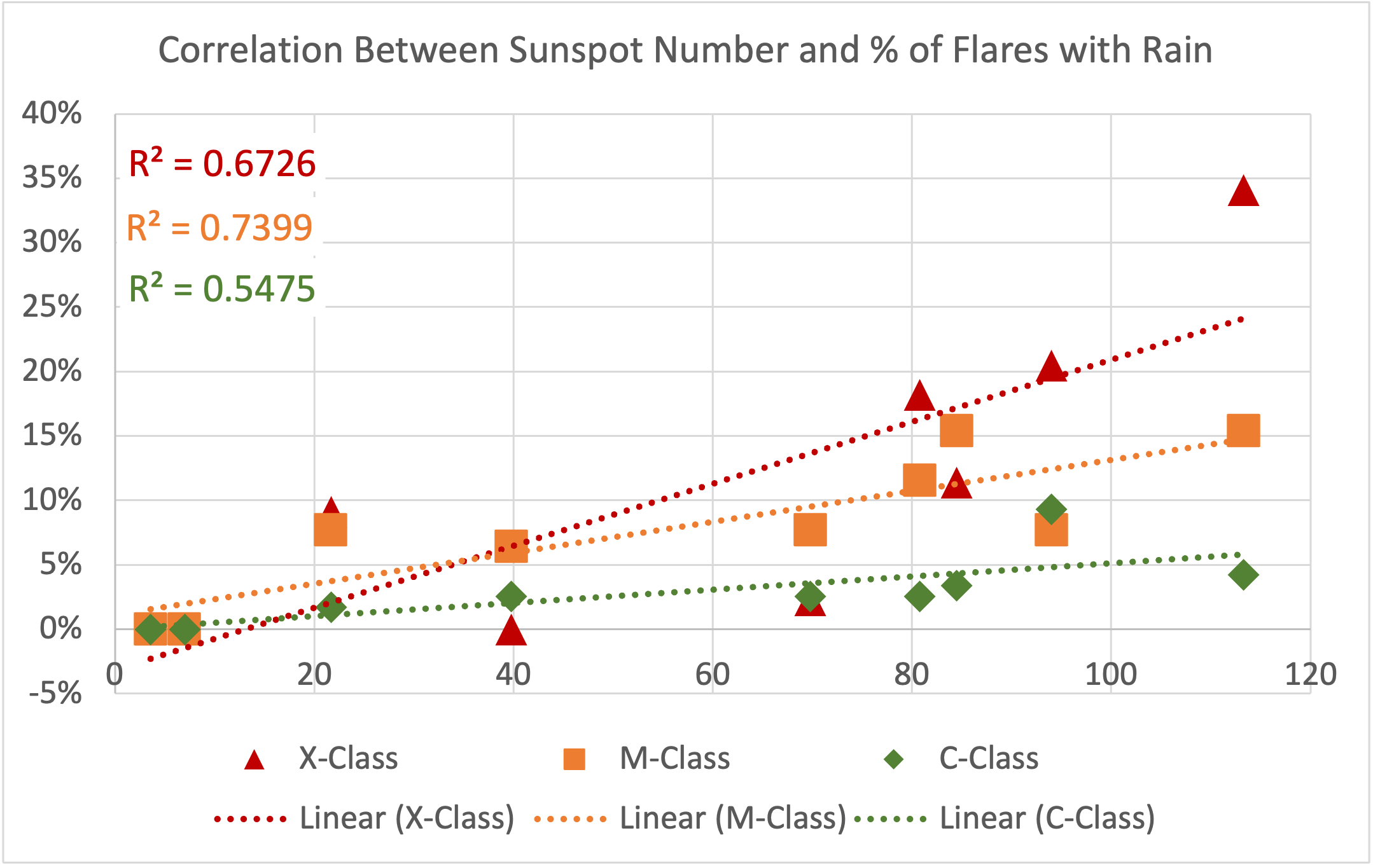}
   \includegraphics[width=0.5\textwidth,clip=]{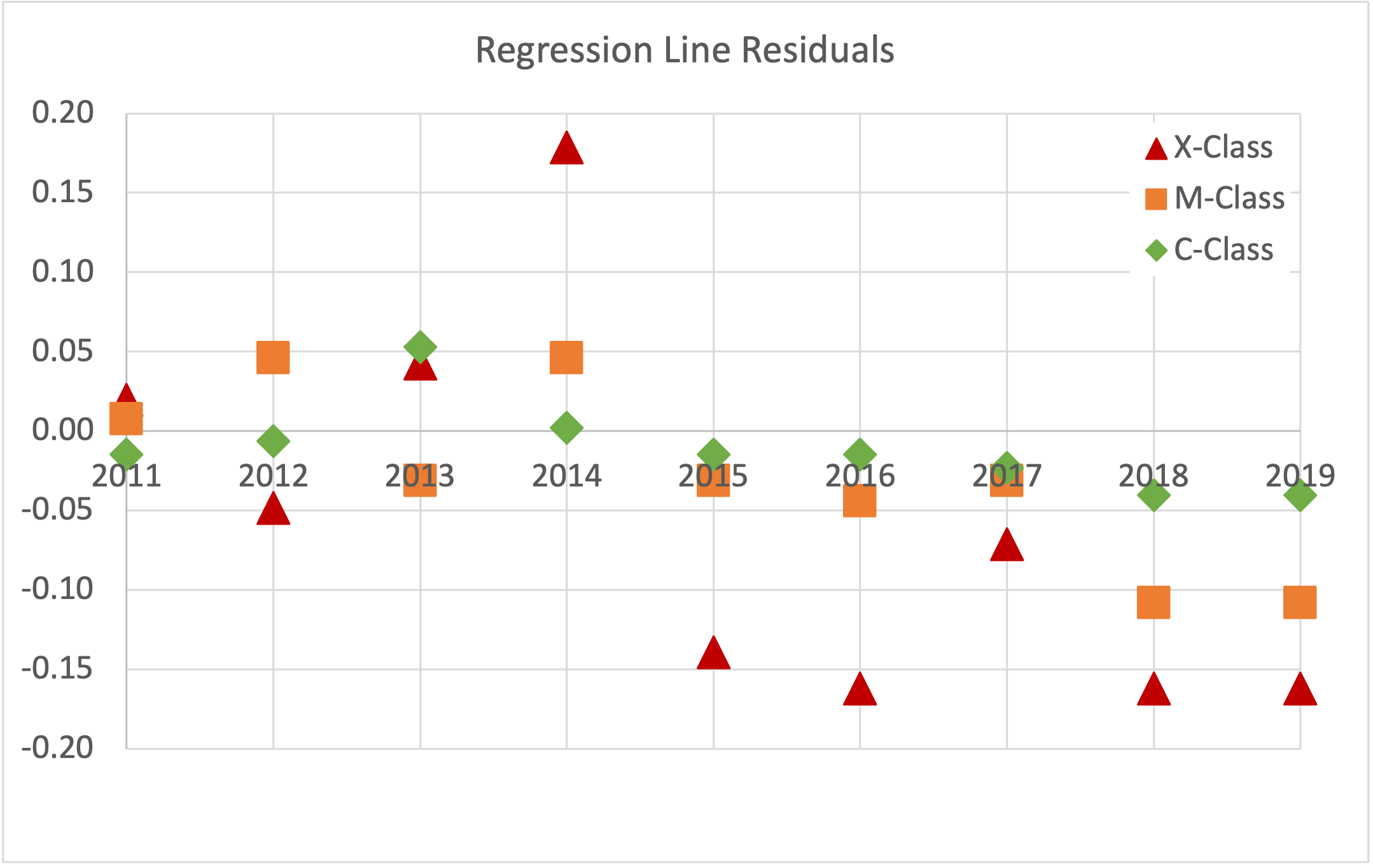}}
  \caption{Top: This graph shows the relationship between the proportion of post-flare arcade rain occurrences (by GOES class) and the solar cycle as plotted by the annual sunspot number. Bottom left: Sunspot number versus the percentage of flares in each class that produce rain. Linear fit lines have been plotted and their respective $r^2$ values are shown on the left side of the graph. All of are reasonably good fits, ranging from about 0.55 (C-class) to 0.74 (M-class). These may prove to be useful tracers for the solar cycle. Bottom left: This graph shows the residuals for all three classes of flare; the distribution of all three are random, showing that the linear fit is a good choice for these relationships.}
   \label{F7}
   \end{figure}

Next, we explored a potential connection between the eruptivity of a flare and whether it produced post-flare arcade rain, since both are well-known to correlate separately with GOES class. Figure \ref{F5} shows a nested pie chart for all three classes of flares, indicating the proportions of the flares which fell into the classifications CME/post-flare arcade rain, CME/no post-flare arcade rain, no CME/post-flare arcade rain, and no CME/no post-flare arcade rain. As has already been well-established by \cite{Kawabata2018} and references therein, most X (here $80\%$) and M-class flares ($71\%$) are eruptive; however, only $13\%$ of the C-class flares were associated with CMEs.

To try to establish whether there was a correlation between CMEs and post-flare arcade rain events, we compared the proportions of post-flare arcade rain/CME and no post-flare arcade rain/CME with no post-flare arcade rain/no CME and post-flare arcade rain/no CME. The results of this are found in Table \ref{T2}. There is a statistically insignificant correlation between eruptivity and post-flare arcade rain occurrence for C-class flares (Fisher's exact test gives a value of 0.204). M-class flares show a nearly statistically-significant relationship between rain presence and eruptivity, with a Fisher test value of 0.084; X-class flares have the poorest relationship, with a Fisher test value of 0.704. We chose the Fisher exact test due to the proportionately small number of non-eruptive flares in X- and M-classes in general, but would still caution against drawing definitive conclusions given the low numbers in the non-eruptive category. While intriguing, further study would be required to show that these relationships are more than simple correlations.

We also compared the Mt. Wilson Hale class of each source active region across the flare classes and against the flares' eruptivity. The results of this investigation are shown in Figure \ref{F6}; there are a few flares of each class in $\alpha$ type regions, but the overwhelming number of flares of all types occur in the more complex $\beta$ designation -- particularly $\beta\gamma\delta$ spots, which are commonly known to be the most active (\cite{Benz2017} and references therein). We saw very similar results when the Hale class was graphed against flare eruptivity, seen in the right graph of Figure \ref{F6}.

Finally, to determine whether the presence of post-flare rain followed the solar cycle, we plotted the post-flare arcade rain-positive flares in each class by year as a percent of all flares of that class, against the SILSO annual sunspot number, obtained from the Royal Observatory of Belgium, Brussels (\url{https://wwwbis.sidc.be/silso/datafiles}). The results can be seen in Figure \ref{F7}. All three classes of flares track the solar cycle in rough terms, with reasonably good fits for positive linear relationships to the solar cycle, as shown in the bottom left graph of \ref{F7}. X-class flares have an $r^2$ value of 0.67, while M- and C-class flares exhibit values of 0.74 and 0.55, respectively. The bottom right graph shows that the residuals for each class are fairly randomly distributed, which means that the linear fit is a reasonable one for this relationship. Therefore, we conclude that X- and M-class flares show promise as indicators of solar cycle strength. While the span covered here did not quite capture a full solar cycle, these data show moderate levels of post-flare arcade rain before the peak, increasing peaks that match the sunspot number peak, and then dying away almost entirely across the years of solar minimum. We discuss this finding in more detail in the next section.

\setlength{\tabcolsep}{10pt}
\begin{center}
\begin{table}[ht]
 \caption{Post-Flare Arcade Rain and CME Occurrence}
 \begin{tabular}{||c c c | c c ||} 
 \hline
 & \multicolumn{2}{c|}{Post-Flare Arcade Rain} & \multicolumn{2}{c||}{No Post-Flare Arcade Rain}\\
 \hline
 & \textit{No CME} & \textit{CME} & \textit{No CME} & \textit{CME}\\ [0.5ex] 
 \hline
 \textit{X-class} & 7 & 35 & 0 & 2 \\ [0.5ex]
 \textit{M-class} & 14 & 42 & 9 & 13 \\ [0.5ex] 
 \textit{C-class} & 26 & 5 & 76 & 11 \\ [0.5ex] 
  \hline
 \end{tabular}
 \label{T2}
\end{table}
\end{center}
\section{Implications}\label{s:imps}

The purpose of this paper is to establish a benchmark set of statistics for a well-known phenomenon which has nevertheless been under-reported. The main findings of this statistical study can be summarized as follows:

\begin{enumerate}
    \item \textbf{Post-flare arcade rain is prevalent to the point of ubiquity in X-class flares \citep[as previously reported, see ][]{Benz2017}, and very prevalent across the M-class flare population. However, flare duration and post-flare arcade rain duration are \emph{not} well-correlated.} One notable subset we identified, confined jets, did not have identifiable post-flare arcade rain despite relatively strong flares. We suspect that this is a simple matter of insufficient spatial resolution, since there is plenty of work showing condensations in larger jets with the same dynamics (\cite{Raouafi2016}, \cite{Kumar2019}, \cite{Mason2019}). There is also a distinct, direct relationship between GOES class and post-flare arcade rain duration, suggesting that more sets of loops reconnect across the neutral line to form arcades in stronger flares. Post-flare arcade rain occurs, though with markedly reduced frequency, in C-class flares, making up only approximately half of those analyzed.
    \item \textbf{X- and M-class post-flare arcade rain may be a useful tracker of the solar cycle.} While rain in all flares tracked the solar cycle to a moderate extent, the highest correlation occurred for M-class flares, followed by X-class flares. If this correlation is shown to extend across more than one solar cycle, the rate of raining X- and M-class flares may be useful at the early stages of a solar cycle to predict the strength of solar maximum, and later in a cycle to tell how quickly solar minimum is approaching (with the caveat that weaker flares are more common, particularly as active regions weaken near the tail of the cycle). Further research is planned to corroborate this finding.
    \item \textbf{The presence or absence of post-flare rain is not consistently correlated with whether the flare was eruptive or not.} As pointed out in the discussion concerning Table \ref{T2}, X- and C-class flares do not show any relationship between eruptivity and rain presence; M-class flares do, although it is not quite statistically significant. More work is required to settle whether the weak correlation is truly due to the post-flare arcade rain or simply the fact that the majority of X and M-class flares are eruptive and the majority of C-class flares are not, producing significantly smaller values in one category in each case. Regardless, this lack of strong, consistent correlation is perhaps unsurprising; while the two are connected by the process of reconnection, which influences both the creation of the CME and the amount of energy injected into the arcade loops, the dynamics which drive the arcade after its formation (radiative cooling and conduction) are physically isolated from and occur at different time scales from those driving CME creation (magnetic reconnection). By all accounts, flare reconnection occurs in a very short time period. Once the reconnection has happened, the loop cools in relative isolation, disconnected from the presence or absence of a CME evolving above it. Certainly, if one considers the pool of all flares, the strongest are more likely to host CMEs and more likely to host post-flare arcade rain, and our findings support this. Within a class, though, the reconnection should be roughly comparable across all of the flares, and there we do not see a strong correlation between CME and post-flare arcade rain occurrence. It is entirely possible for, as an example, an M-class arcade loop to simply cool non-catastrophically and drain (with a CME or without), and it is also possible for an M-class arcade loop to cool catastrophically and form post-flare arcade rain (with a CME or without).
    \item \textbf{Post-flare rain can act as a proxy for flare energy release.} The near-universal prevalence of post-flare arcade rain in X and M-class flares, coupled with its distinct drop in occurrence around C5, point to a direct energy correlation for condensation formation in arcades. Further research is required to better define the low-end energy cutoff and other factors which may affect it. However, as individual post-flare arcade rain condensation geometry estimates have been carried out before (\cite{Antolin2015}), and radiative cooling rates are well-established, these observations of post-flare arcade rain presence and duration can be utilized as a proxy to aid in calculating well-constrained estimates of the energy released during impulsive-phase flare reconnection.  
    \item \textbf{Post-flare arcades can persist for days after the flare, eventually transitioning from post-flare rain to quiescent active region condensations.} While most C-class flares did not produce arcades, and most X-class flares produced arcades that faded after a few hours, many M-class and some X-class flares produced arcades that continued to be visible in 171 Å for several days after the flare, well after the ribbons had ceased evolving. At this point, the pronounced region of higher emission at the loops' tops dissipated, and the loops proceeded to produce more erratic condensations that were no longer coordinated with neighboring loops. This shift was subtle but common, and points to a shift from post-reconnection condensation formation to the ubiquitous quiescent active region thermal nonequilibrium condensations that have been widely discussed. To the authors' knowledge, this has not been previously reported; the life span of such loops, which do not appear to reconnect again readily after the flare, may be useful for studying active region aging.
\end{enumerate}

\section{Acknowledgements}
The authors would like to thank the reviewer for constructive suggestions that greatly improved the manuscript, and Jeff Reep for suggesting this paper topic several years ago. EIM was supported during the paper's early development by an appointment to the NASA Postdoctoral Program at the Goddard Space Flight Center, administered by Universities Space Research Association under contract with NASA, and later by NSF's Solar-Terrestrial Program (Award No. AGS-1923377). KK contributed to the research described in this paper through an internship arranged at the Goddard Space Flight Center through the Naval Academy's Summer Internship Program. Both authors gratefully acknowledge collaboration with the NASA GSFC Internal Scientist Funding Model program ``Magnetic Energy Buildup and Explosive Release in the Solar Atmosphere".

\bibliography{solar_rain_stats}{}
\bibliographystyle{aasjournal}

\end{document}